\documentclass{tlp}
\pdfoutput=1
\usepackage{ifthen}
\usepackage{url}
\usepackage{swipl}
\usepackage{amssymb}

\usepackage{color}

\submitted {5th October 2009}
\revised {24th February 2010}
\accepted{22nd November 2010}

\usepackage[pdftex]{graphicx}
\DeclareGraphicsExtensions{.pdf,.jpg,.png}
\graphicspath{{figs/}{./}}

\newcommand{\fnurl}[1]{\footnote{\url{#1}}}
\begin{document}
\bibliographystyle{acmtrans}

\long\def\comment#1{}

\title{SWI-Prolog}

\author[J. Wielemaker et al.]
{JAN WIELEMAKER \\
VU University Amsterdam\\
\email{J.Wielemaker@cs.vu.nl}
\and
TOM SCHRIJVERS \\
Universiteit Gent \\
\email{Tom.Schrijvers@ugent.be}
\and
MARKUS TRISKA \\
Technische Universit\"at Wien \\
\email{triska@dbai.tuwien.ac.at}
\and
TORBJ\"ORN LAGER \\
University of Gothenburg\\
\email{lager@ling.gu.se}
}

\pagerange{\pageref{firstpage}--\pageref{lastpage}}
\volume{\textbf{?} (?):}
\setcounter{page}{1}

\maketitle

\label{firstpage}

\begin{abstract}
SWI-Prolog is neither a commercial Prolog system nor a purely academic
enterprise, but increasingly a community project. The core system has
been shaped to its current form while being used as a tool for building
research prototypes, primarily for \textit{knowledge-intensive} and
\textit{interactive} systems. Community contributions have added
several interfaces and the constraint (CLP) libraries. Commercial
involvement has created the initial garbage collector, added several
interfaces and two development tools: PlDoc (a literate programming
documentation system) and PlUnit (a unit testing environment).

In this article we present SWI-Prolog as an integrating tool, supporting
a wide range of ideas developed in the Prolog community and acting as
glue between \textit{foreign} resources. This article itself is the
glue between technical articles on SWI-Prolog, providing context and
experience in applying them over a longer period.

To appear in Theory and Practice of Logic Programming (TPLP)
\end{abstract}

\begin{keywords}
Prolog, logic programming system
\end{keywords}


\section{Introduction}
\label{sec:intro}

SWI-Prolog was started as a recreational program for personal
understanding and enjoyment after reading \citeN{Bowen:83}. This toy
landed in fertile soil. Our university department was involved in
Shelley \cite{Anjewierden:90b} a workbench for knowledge engineering
that was to be implemented using Quintus Prolog (version 1) and PCE, an
in-house developed object-oriented graphical library written in C.
Frustrated by the inability of Quintus Prolog version 1 to make
recursive calls between C and Prolog, which seriously complicated our
integration between Prolog and the graphical system, we demonstrated the
much cleaner integration achievable with the small SWI-Prolog
system.

While SWI-Prolog was clearly inferior to Quintus in terms of robustness
and execution speed, it quickly replaced Quintus, not only at our site,
but also at the two overseas sites where modules for the workbench were
being developed. Why did this happen? We think this was due to three
small, yet much valued features: (1) the \index{make/0}\predref{make}{0} predicate that reloads
all files that have been edited since last loaded (see
\secref{editcycle}); (2) the ability to run \index{make/0}\predref{make}{0} on a restarted
saved state dramatically reduced the application restart-time; (3) using
the \emph{auto-loader} the initial image could be restricted to the core
parts of the application, which again reduced memory usage and startup
times. The faster and bi-directional interface to the graphics libraries
made the application much more responsive for typical interaction.

Because the market for commercial Prolog systems was by then (late 80s)
already crowded, we decided to make the system available for free to the
community through anonymous ftp. Initially, this did not create a
\emph{developer} community, but it did create a \emph{user} community.
The user community consisted mostly of universities that used this
small, free and portable Prolog system for education. This unforeseen
development was strengthened when SWI-Prolog was ported to MS-Windows as
part of another research project.

SWI-Prolog's development has been guided by internal projects, external
(including commercial) users and increasingly by developers from the
community. This article summarizes the distinguishing features. Because
most of the technical details are dealt with in other articles, we
concentrate on providing an overview, motivating our decisions and
describing our experiences.

In \secref{appcontext} we describe the development of SWI-Prolog, with a
particular emphasis on how it was embedded in research projects. The
paper continues with a user-oriented description, covering the
environment (\secref{environment}), the constraint libraries
(\secref{clp}), interfaces to the outside world (\secref{interfaces})
and finally web-applications (\secref{web}). The next part of the
paper targets the Prolog developers community and addresses the language
properties and regression testing. We end with a brief discussion and
a description of future work.

\section{Developing Prolog in the context of applications}
\label{sec:appcontext}

In the 90s, before SWI-Prolog attracted a wider network of developers, it took
form in the SWI department at the University of Amsterdam. There it was used as
an in-house (or \jargon{in-project}) tool for the development of
proof-of-concept software prototypes, rather than an objective on its own
right. This supporting role has very much influenced both the development
process and design decisions of SWI-Prolog. New architectures, features and
deployment strategies for SWI-Prolog were explored in the course of new
research projects and as demanded by every new prototype. Lessons learned
influenced subsequent extensions of the core and its libraries.

\subsection{Using Prolog as glue}
\label{sec:uniform}

SWI-Prolog's supporting role in the academic research projects is primarily a
tool for rapid development. It provides a uniform programming environment for
accessing a range of resources. Using multiple environments requires
interfaces that are generally hard to program due to differences in datatypes,
control (e.g., Prolog non-determinism) and organization (e.g., object-oriented
vs.\ functional).

The overall approach we have followed over the years is to bring the required
resources to Prolog, either as Prolog resources (e.g., our HTTP client and
server libraries) or as encapsulated foreign resources (e.g., our RDF store,
\citeN{Wielemaker:03a}).
The latter makes up for the wealth of resources that are not natively available
in Prolog, either for lack of development time or because Prolog is not a
suitable language (e.g., libraries that require significant destructively
updatable state such as a graphics library).

In general, the way to deal with the complexity of an environment that
offers many different resources is to define stable and, preferably,
small interfaces. While, in our opinion, this is a boon in large
top-down designed software systems, it does form a significant burden
for developing the medium-scale applications (e.g., 50,000 lines) that
we target. Our projects consist of small teams (2 to 7 programmers) that
quickly evolve interfaces and incorporate new insights.

\subsection{Application-Driven Requirements}

SWI-Prolog's supporting role in the application-oriented research projects
helped setting its main requirements:

\begin{itemize}
    \item In its role as uniform platform, the system must be able to
     encapsulate foreign resources in a flexible and transparent way.

    \item To support rapid and incremental development, the system must
    load large programs rapidly. After editing, a program must be synchronized
    quickly without losing state stored in dynamic predicates and foreign
    resources.

    \item Decent tools are essential to facilitate (rapid) development.
    In particular, it provides good editor support, a source-level debugger, an
    execution profiler and a cross-referencer.

    \item In order to qualify as a web-based application platform, SWI-Prolog
    must run as a server, $24\times 7$.  As a server, it must be stable and free of
    resource leaks. Moreover, to provide scalable request handling, it must exploit
    concurrent (multi-core) hardware.

    \item Reflexiveness is desirable for application-specific
    program analysis and transformation, and supports the debugger.
    Regardless of the compilation mode, good debugger support and the ability
    to inspect code (e.g., \index{listing/1}\predref{listing}{1}) should be provided .

    \item In order to support all major desktop and server
    systems for applications, SWI-Prolog should be portable.
    Currently, SWI-Prolog is portable across platforms that provide a
    C99-compliant C-compiler and implement either the POSIX/X11 or the
    Win32/Win64 APIs.
\end{itemize}

\paragraph{Main Application Influences}

The directly supported research prototypes were (and in part still are)
interactive applications for the management of knowledge models through
graphical user interfaces. This has prominently resulted in both the XPCE
library and the early adoption of RDF.
\begin{itemize}
\item
XPCE \cite{DBLP:conf/lpe/WielemakerA02} provides
an---for those days---advanced and tightly integrated object-oriented
framework for the development of graphical user interfaces.
\item
RDF (Resource Description Format, \citeN{Lassila:99a}) provides a widely
accepted and extensible infrastructure for representing knowledge.
RDF also facilitates exchange of knowledge-bases between SWI-Prolog
and external tools.
\end{itemize}

Support for networking and subsequently concurrency started when we used Prolog
programs as an ``intelligent agent'' in a FIPA-based agent framework
\cite{DBLP:journals/spe/BellifeminePR01}. After a while, we realized that our
job would become much easier if Prolog also provided support for concurrency.
Support for HTML was added when we used Prolog
to partition text into logical segments and classify these segments with
terms from an ontology. It also turned out that full support for Unicode
is necessary to deal with character-entities in HTML.

Both HTML and HTTP support went through a number of iterations, where
the code was part of projects, but not of the SWI-Prolog libraries.
The core infrastructure for these and many other extensions reached the system
libraries in the MultimediaN project around 2005 (see \secref{web}).

Often one thing leads to another. To support good performance for arbitrary
reasoning patterns, our RDF infrastructure has to be main-memory based. An
unfortunate consequence is that applications with significant amounts of data
take relatively long to start. In order to use such applications effectively as
web services and avoid a prohibitive start-up cost at every request, the
application has to be able to run continuously. Moreover, to promptly serve
simultaneous clients, such applications must also be concurrent.
\section{The SWI-Prolog development environment}
\label{sec:environment}

The development environment is a crucial part of a Prolog system that
aims at prototyping large applications. SWI-Prolog's user-friendliness
stems from three sources: (1) command-line interaction, (2) graphical
tools, and (3) design decisions for the compiler and extensions
to the language. The last category is described in \secref{language}. In
this section we take a closer look at the command-line and graphical
tools.


\subsection{Prolog top-level interaction}
\label{sec:toplevel}

Originally, the top-level interaction of SWI-Prolog was based on the
Edinburgh tradition, prompting for alternatives if and only if the query
contains variables and printing \textit{Yes} or \textit{No}, the only
small difference being that user replies were processed on
single-keystrokes (i.e., without using ``return''). This approach
suffers from three problems: (1) the top-level syntax was not suitable
for copy/paste into the top level, (2) there is no way to deal with
non-deterministic goals that have no variables, and (3) there is no
clean way to represent residual constraints. We have revised the
SWI-Prolog top level based on one simple principle: ``The answer
substitution is a valid Prolog goal that returns the same answers as the
original query''. Starting from this principle, the rest follows
naturally. \Figref{toplevel} illustrates some typical cases.

\begin{itemize}
    \item An answer substitution is a conjunction of equalities of the
          form \arg{Var} = \arg{Value}. If there are no variables,
	  the answer is simply \const{true}. (1 in \figref{toplevel}).
    \item If variables in the answer carry constraints, \index{copy_term/3}\predref{copy_term}{3}
	  is used to create a copy without constraints and goals to
	  reinstate the constraints.  These goals are printed after
	  the variable-bindings. (4 in \figref{toplevel}).
    \item A query that succeeds deterministically writes its answer
          substitution followed by a full-stop and prompts immediately
	  for the next query. (2 in \figref{toplevel}).
    \item A query that fails writes \const{false.}  The predicate \index{false/0}\predref{false}{0}
    	  is a built-in.
    \item A query that succeeds non-deterministically waits at the end
	  of the printed answer substitution.  If the user types `;',
	  this is echoed and the system returns the next substitution.
	  (3 in \figref{toplevel}). If the user hits \textsc{return},
	  the system prints a full-stop.
\end{itemize}

\begin{figure}
\begin{code}
1 ?- [library(clpfd)].
true.                   

2 ?- A is 1+1.
A = 2.                  

3 ?- member(x, [a,x,y]).
true ;                  
false.

4 ?- A #> 10.
A in 11..sup.           
\end{code}

\noindent
    \caption{Top level interaction in SWI-Prolog}
    \label{fig:toplevel}
\end{figure}

\subsubsection{Command line editing}		\label{sec:cmdedit}

During system development, developers spend a considerable amount of
time entering commands, specially writing test-queries to assess correctness
for parts of the application being developed. SWI-Prolog provides the
following features to support this development mode:

\begin{itemize}
    \litem{Using (GNU-)readline for the top level input}
\emph{Completion} of the library is extended with completion on
alpha-numerical atoms which enables faster input of long predicate
identifiers and atomic arguments, as well as inspection of the possible
alternatives using Alt-?. The completion algorithm uses the built-in
completion of file names if no atom matches.

    \litem{Command line history}
SWI-Prolog provides a history facility resembling the
corresponding facilities in the Unix \program{csh} and \program{bash}
shells. Viewing the list of executed commands (using \verb$?- h.$) is a
particularly valuable feature.

    \litem{Top-level bindings}
The facility to reuse answer substitutions through copy/paste is useful,
but limited to bindings that have been printed recently, have not been
modified using the \index{portray/1}\predref{portray}{1} hook and are short. For this reason
SWI-Prolog stores the variable-bindings from top level queries in the
database under the name of the used variable. Top level query expansion
replaces terms of the form \$\arg{Var} (\$ is a prefix operator) with
the last recorded binding for this variable. New bindings due to
backtracking or new queries overwrite the old value.  Typical example:
by using \verb'$X' the user avoids typing or copy/paste of the object
reference returned by a call to XPCE:

\begin{code}
1 ?- new(X, picture).
X = @12946012/picture.  

2 ?- send($X, open).
true.
\end{code}

\noindent
\end{itemize}

\subsection{Supporting the edit cycle}
\label{sec:editcycle}

There are two simple but frequent tasks involved in the edit-reload
cycle: finding the proper source, and reloading the modified source
files. SWI-Prolog supports these tasks with two predicates:

\begin{description}
    \predicate{make}{0}{}
SWI-Prolog maintains a database with information about every loaded
file: the pathname of the file, the time of the most recent modification
of the file (time stamp) that was valid when the file was loaded, and
the context module from which it was loaded. The \index{make/0}\predref{make}{0} predicate checks
whether the modification time of any of the loaded files has changed and
reloads these file into the proper module contexts. After updating the
running program, \index{make/0}\predref{make}{0} lists undefined predicates as described in
\secref{check}.

    \predicate{edit}{1}{+Specifier}
Finds all entities with their location that match \emph{specifier}. If
there are multiple entities related to different source files asks the
user for the desired one and calls the user-defined editor, placing the
cursor at the location of the selected entiry. The predicate searches
for (loaded) files, predicates and modules. The interface can be
customized in two ways: by extending the entities searched for (e.g.,
XPCE classes, see \secref{interfaces}), and by changing the editor that
is called. Below is an example:

\begin{code}
?- edit(rdf_tree).
Please select item to edit:

  1 class(rdf_tree)             'rdf_tree.pl':27
  2 module(rdf_tree)            'rules.pl':460

Your choice? 2
\end{code}

\noindent
\end{description}

\subsubsection{DWIM: Do What I Mean}		\label{sec:dwim}

DWIM (\emph{Do What I Mean}) is implemented at the top level to quickly
fix mistakes and allow for under-specified queries. It corrects the
following errors: simple spelling errors, different word-order (e.g.,
\emph{exists_file} matches \emph{file_exists}), arity mismatches and
wrong module.

DWIM is used in three areas. First, queries typed at the top level are
checked, and if there is a unique correction the system prompts the user
whether the corrected query is to be executed instead of the original
one. Especially adding the module specifier improves interaction from
the top level when using modules. If there is no unique correction the
system reports all close candidates. Second, Predicates such as \index{spy/1}\predref{spy}{1}
act on the named predicate in any module if the module is omitted.
Third, if a predicate existence error is not caught, the DWIM
system is activated to report likely candidates.

\subsection{Quick consistency check}
\label{sec:check}

The library \emph{check} provides quick tests on the completeness
of the loaded program. The predicate \index{list_undefined/0}\predref{list_undefined}{0} searches the
internal database for predicate structures that are undefined (i.e.,
have no clauses and are not defined as dynamic or multifile).  Such
structures are created by the compiler for a call to a predicate that
is not yet defined.  In addition, the system provides a primitive that
returns the predicates referenced from a clause by examining the
compiled code.  \Figref{listundef} shows the partial output of running
\index{list_undefined/0}\predref{list_undefined}{0} on the \emph{chat 80} \cite{PereiraShieber87} program.

\begin{figure}[ht]
\begin{code}
1 ?- [library(chat)].
true.

2 ?- list_undefined.
Warning: The predicates below are not defined. If these are defined
Warning: at runtime using assert/1, use :- dynamic Name/Arity.
Warning:
Warning: chat:standard/4, which is referenced by
Warning:        chat:i_adj/9 at /home/jan/lib/prolog/chat/slots.pl:128
...
\end{code}

\noindent
    \caption{Using \index{list_undefined/0}\predref{list_undefined}{0} on chat 80 wrapped into the module
	     \const{chat}. To save space only the first of the 9
	     reported warnings is included.}
    \label{fig:listundef}
\end{figure}

\subsection{Printing log messages}

The library \emph{debug} is a lightweight infrastructure that handles
printing debugging messages (logging) and assertions. Each debug message
is associated with a \arg{Topic}. A \arg{Topic} is an arbitrary Prolog
term that identifies a class of debug messages. Using compound terms
such as \verb$http(connection)$ and \verb$http(query)$, all debugging
messages related to HTTP can be enabled with \verb$?- debug(http(_))$.

\begin{description}
    \predicate{debug}{3}{+Topic, +Format, +Arguments}
Prints a message using \term{format}{Format, Arguments} if \arg{Topic}
unifies with a topic enabled with \index{debug/1}\predref{debug}{1}.

    \predicate{debug/nodebug}{1}{+Topic [$>$file]}
Enables/disables messages for which \arg{Topic} unifies. If
\verb$>$\arg{file} is added, the debug messages are appended to the
given file.

    \predicate{assertion}{1}{:Goal}
Assumes that \arg{Goal} is true. Prints a stack-dump and traps
to the debugger otherwise. This facility is derived from the
\funcref{assert}{} macro as used in C, renamed for obvious reasons.
\end{description}

Calls to \index{debug/3}\predref{debug}{3} and \index{assertion/1}\predref{assertion}{1} are replaced with \const{true} using
\jargon{goal-expansion} if optimization is enabled.


\subsection{The built-in editor}
\label{sec:pceemacs}

\emph{PceEmacs} is an Emacs clone written in XPCE/Prolog. It has full
access to the application by means of the reflexive capabilities of
Prolog. On each key-stroke, Prolog opens the edit buffer as a stream and
tries to read the current clause. If there is a syntax error, it
displays unobtrusive information about the location. If the syntax is valid,
the clause is colored based on information from the latest
cross-reference analysis. Goals are given a menu that provides access to
the source, documentation, and listing. Singleton variables are
highlighted. If the cursor appears inside the name of a variable, all
other occurrences of this variable in the clause are underlined.
Whenever the user pauses for two seconds, Prolog opens the edit buffer
as a stream and performs a full cross-reference of the edit buffer.
\Figref{guitracer} shows PceEmacs embedded in the debugger.

\subsection{The source-level debugger}
\label{sec:guitracer}

The hook-predicate \term{prolog_trace_interception}{+Port, +Frame,
+Choice, -Action} can be implemented to realize an alternative
debugger such as the source-level debugger described below. The
source-level debugger provides three views (\figref{guitracer}):

\begin{itemize}
    \litem{The source}
An embedded \emph{PceEmacs} (see \secref{pceemacs}) shows the current
location, indicating the current port using color and icons.
\emph{PceEmacs} also allows setting a \emph{breakpoint} on an arbitrary
location in a clause.  Breakpoints are realized by replacing a
virtual machine instruction with a \emph{break} instruction which traps
to the debugger, finds the instruction it replaces in a table and
executes this instruction.

    \litem{Variables}
The debugger displays a list of variables appearing in the current
frame, with their names and current bindings in the top-left window (see
\secref{compiler}). The representation of values can be changed using
the familiar \index{portray/1}\predref{portray}{1} hook. Double-clicking the displayed value of
a variable opens a separate window showing the variable binding with
additional layout to clarify the structure of a term.

    \litem{The stack}
The top-right window shows the recursion stack as well as the recent
outstanding choicepoints. Any node can be selected to examine the
context of that node. The stack window allows one to quickly examine
choicepoints left after a goal succeeded.  Clicking a choice-point
shows the clause that last succeeded.  Using the \jargon{up} command
shows the source of the calling context.
\end{itemize}

\postscriptfig[width=0.8\linewidth]{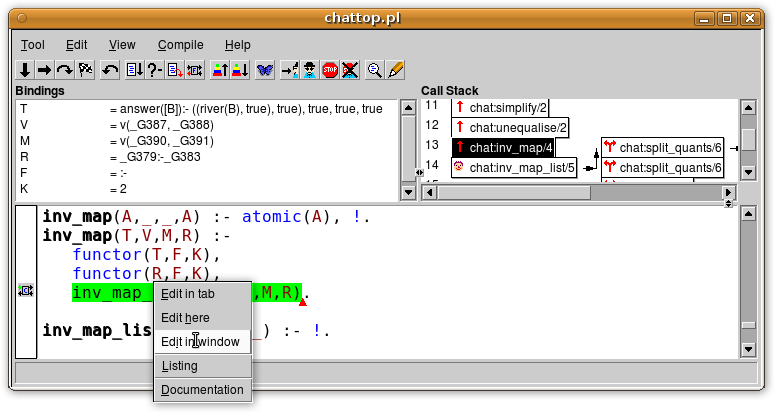}{The Source-level
Debugger. The source code is rendered using an embedded version of
PceEmacs.}

\subsection{Execution Profiler}			\label{sec:prof}

The \emph{Execution Profiler} builds a call-tree at runtime and counts
the number of calls and redos for each node in this call-tree. The time
spent in each node is established using stochastic sampling. Prolog
primitives are provided to extract all information from the recorded
call-tree. A graphical Prolog profiling tool presents the information
interactively, similarly to the GNU \program{gprof} \cite{graham82gprof}
tool (see \figref{prof}).

\postscriptfig[width=0.8\linewidth]{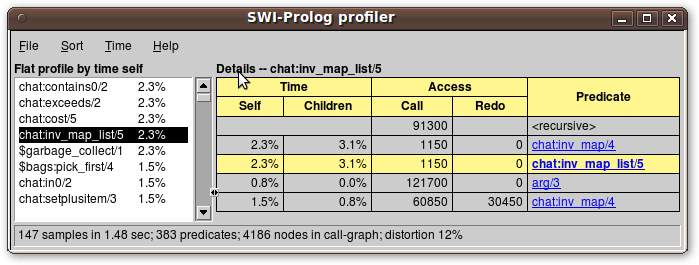}{The Profiler showing
information about CHAT80.  To profile a goal, run it using
\texttt{?- profile(Goal).}}

\subsection{Graphical cross-referencer}
\label{sec:gxref}

\Figref{gxref} shows the output of running \index{gxref/0}\predref{gxref}{0}, which shows the
dependencies between source files based on cross-reference analysis for
all files loaded into the running system. The analysis is provided by a
separate public library called \file{prolog_xref.pl}, which is also used
by PceEmacs. The (red) exclamation-mark indicates that there is at least
one warning in the file or directory. In addition to the functionality
exemplified by \figref{gxref}, the tool can show the dependencies
between sources as a graph and it can generate module declarations to
help transforming non-modular code into modular code.

\postscriptfig[width=0.9\linewidth]{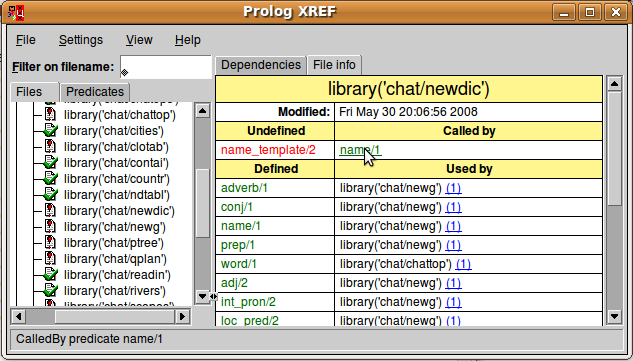}{The cross-referencer}

\subsection{Discussion}

Many of the current tools are built in SWI-Prolog's proprietary graphics
system (XPCE), though the underlying computation for coloring,
cross-referencing, profiling, and tracing is accessible from Prolog
through public APIs. In an ideal world, these tools would be neatly
integrated in open IDEs such as Eclipse. In practice, such an
integration is hard to achieve. \citeN{DBLP:journals/corr/abs-0903-2252}
have implemented a Prolog parser in Java for this purpose. Still, the
result may be incorrect, e.g., because the IDE does not know which
operators are visible.
\section{Constraint libraries}
\label{sec:clp}

Constraint Logic Programming functionality came rather late in the
lifetime of SWI-Prolog, because it lacked the basic support. This
changed early in 2004 when \jargon{attributed variables} were added to
the language (see \Secref{language:attvars}). The Leuven CHR library was
then the first CLP library to be ported to SWI-Prolog. Later came a port
of Christian Holzbaur's CLP($\mathbb{QR}$) library, and a finite domain
CLP(FD) solver. Finally, we mention SWI-Prolog's INCLP($\mathbb{R}$)
library \cite{inclpr}, which provides non-linear constraints over
the reals, and was implemented on top of CHR.

\subsection{CHR}

The Constraint Handling Rules (CHR) language was created about 20 years
ago \cite{fruUNchrUNoverviewUNjlp98}. Since then CHR has proved its
merit as a powerful complement to Prolog. Both have a firm basis in
logic, but whereas Prolog is about single-headed rules, backward
chaining and backtracking, CHR has multi-headed rules, forward chaining
and committed choice.

CHR's features support the modeling and implementation of constraint
solvers. CHR has also turned out to be very useful for applications of
(Term) Rewriting Systems and Production Rule Systems, as well as for
expressing imperative algorithms in a high-level manner.

CHR is usually embedded in Prolog as an add-on library and for a long
time the SICStus implementation by Christian Holzbaur
\cite{holzUNfruUNprologUNchrUNcompilerUNaai00} has been the standard
implementation. This library was also available in YAP
\cite{DBLP:conf/iclp/SilvaC07}; two different, older implementations
come with ECLiPSe \cite{DBLP:conf/padl/WallaceS99}. Neither of these
constraint systems was under active development in the last decade. In
the last five years, the K.U.Leuven CHR system
\cite{schrUNdemoenUNkulchrUNchr04} has come to replace these older
systems.

\subsubsection{The K.U.Leuven CHR system}

The K.U.Leuven CHR system started out as small practical project, to
obtain a benchmark for Bart Demoen's new implementation of dynamic
attributed variables for hProlog \cite{Demoen:CW350}. The output of
SICStus' CHR was reverse engineered to obtain a base system. It
became clear very soon that there was much potential for improving the
system, and gradually the system diverged from its roots. It became the
central topic of Tom Schrijvers' Ph.D.\ thesis \cite{schrUNphdthesis05},
and the starting point for subsequent theses by Jon Sneyers, Leslie De
Koninck and Peter Van Weert.

In July 2004 the system was ported to XSB and integrated with tabled
execution~\cite{schrUNwarrenUNchrUNxsbUNiclp04}. Half a year later Jan
Wielemaker was looking for an easy way to provide general constraint
solving capabilities to SWI-Prolog. The smoothness of the port to XSB
convinced him that the K.U.Leuven CHR system was what SWI-Prolog needed
\cite{schrUNwielemakerUNdemoenUNchrUNswiUNwclp05}. From then on
SWI-Prolog has been the K.U.Leuven CHR system's main supported platform,
and Jan Wielemaker has taken care of tight integration with the rest of
the system, notably with the debugger. Gradually other open source
Prolog systems became interested in this new CHR system, and today you
can find K.U.Leuven CHR in hProlog, XSB, SWI-Prolog, YAP, Ciao, B-Prolog
and SICStus. Simplifying the porting process of the CHR library remains
one of the key challenges. Standardization of the core language features
in K.U.Leuven CHR would be a good step in this direction.

\subsection{CLP(FD)}

Finite domain constraint solvers are almost a standard component in
modern Prolog environments. SWI-Prolog's solver is implemented
completely in Prolog. We aimed at providing a solver that is very
reliable, rather than exceedingly fast. Users can easily modify and
extend the solver by following a few simple conventions that are
explained in the solver's user manual. Because finite domain constraints
are also used in introductory Prolog courses, we have implemented several
features that make SWI-Prolog's finite domain constraint solver
suitable as a teaching aid for beginners. In the following
subsections, we discuss these features in more detail.

\subsubsection{Extending traditional finite domain solvers}
\label{sec:clpfdintro}

The need for arbitrary precision integer arithmetic is widely
recognized, and many common Prolog systems provide transparent built-in
support for arbitrarily large integers.

It thus seemed natural to enhance a constraint solver over finite
domains with the ability to reason over arbitrarily large integers.
SICStus Prolog already goes in that direction, using the symbolic
constants \const{inf} and \const{sup} to denote default domain
limits, but internally, they still correspond to quite small integers --
the system yields \textit{representation errors} when these limits
are exceeded.

We have implemented a new constraint solver over finite domains, in
which big integers are transparently used. We accept the
\const{inf}/\const{sup} notation of SICStus Prolog, but these
atoms now denote the actual infinities instead of abbreviating
underlying finite limits.

\subsubsection{Ensuring terminating propagation}

By allowing unbounded domains we gain expressibility at the price of
potentially nonterminating propagation. For example, queries like
\verb+X#>abs(X)+ or \verb+X#>Y,Y#>X,X#>=0+ do not terminate in many
existing constraint solvers. \citeN{DBLP:journals/corr/abs-0903-2168}
describe how to guarantee terminating propagation.

\subsubsection{Uniform arithmetic}

Prolog's built-in arithmetic predicates are \textit{moded} -- at
evaluation time, expressions must be ground. Finite domain constraint
solvers remove this restriction. A significant generalization was
achieved in~1996 with the first release of a finite domain system for
SICStus Prolog. The equality relation \verb+#=/2+ could now be used in
place of \index{is/2}\predref{is}{2} for integers. In our solver, we generalize this to
all constraints, such as~\verb+#=</2+ and \verb+#>/2+, which can be used
instead of the built-ins at all places. At compile time, these
constraints are specialized to fall back to moded built-in
arithmetic, reducing the overhead of using CLP(FD) over native
arithmetic.
\section{SWI-Prolog interface libraries}
\label{sec:interfaces}
\label{sec:glue}

As mentioned in \secref{uniform}, we believe that entire applications
should be written in a single language, and that Prolog is well-suited
to the task. To do this, we have to provide support for document
formats, protocols, etc.\ from Prolog. This is opposite to the position
taken, e.g., by the developers of Amzi! Prolog + Logic
Server.\fnurl{http://www.amzi.com} In Amzi!'s view, a logic program is
comparable to a database and accessed from procedural languages: ``Amzi!
moves you toward a unique view of its positioning in the Prolog market.
It aims to be a component of an application written in other
languages.''\footnote{PC AI Review, Sep/Oct 95.}

Using Prolog for what it is good at and embedding it in a conventional
procedural environment has clear advantages, because it does not require
so many developers familiar with Prolog, and makes it possible to
implement a large part of the system in accordance with ``industrial
standards''. However, we believe it is not the most productive approach
for a large class of projects. Accessing Prolog from an imperative
language as a (logical) database engine suffers from what is know as the
``object/relational impedance mismatch'' (\citeNP{ambler:2002,1242682},
\secref{nqueens}). However, Prolog can provide natural APIs to web
document formats (HTML, XML), relational databases and, especially,
schema-less semantic web data (RDF). In addition, embedding Prolog in
traditional procedural languages makes interactive development more
difficult, while careful encapsulation of software developed in other
languages preserves most of the interactive development features.

Our experience shows that embedding Prolog in modern environments (such
as Java or .NET) is particularly painful. Such environments typically
provide threads, automatic memory management using garbage collection
and (in POSIX systems) signal handling. Although the C-interface
(\secref{capi}) does provide primitives to manage Prolog threads, proper
management of resources is far from simple. Creating and destroying an instance
of a Prolog engine for each call to Prolog is generally too expensive. The
API also allows for using a pool of Prolog engines, but allocating
appropriate resources to this pool is a non-trivial problem.
Synchronizing object lifetimes for objects that are referenced from
Prolog is complicated. POSIX defines process-global asynchronous signal
handling, which is used by both JVMs and the Prolog engine. In short, it
is difficult to combine Prolog with such languages within a single
processes. Debugging interface problems is particularly hard.

More promising interaction is achieved by using network-based
communication mechanisms, such as those provided by InterProlog
\cite{DBLP:conf/jelia/Calejo04} or HTTP. However, communication overhead
is then much larger, which limits the usability of such an approach. But
such separation also has significant advantages: it becomes much easier
to isolate and locate problems; moreover, if various services are
provided by separate Prolog threads, one can still carry out traditional
interactive development. See also \secref{web}.

\subsection{Provided interfaces}

This section provides a brief overview of the external interfaces that
are supported by SWI-Prolog. We distinguish two types of interfaces:
document formats that can be read, written and processed (e.g., XML) and
supported protocols (e.g., HTTP).

\begin{description}
    \item[XML, SGML and HTML] (SWI-Prolog and the
			       web, \citeNP{wielemaker:tplp2008})
These core languages of the web are supported through a C parser library
that is also used by XSB Prolog. The library works in two modes: parsing
a document into a ground Prolog term and using call-backs (the
\jargon{event model}). Beside parsing, we provide a library called
\file{html_write.pl} that is used to output HTML. The library provides a
concise and extensible mechanism for producing syntactically correct
(X)HTML, including a modular mechanism for managing required JavaScript
and CSS resources.

    \item[RDF] \cite{Wielemaker:03a}
The RDF support consists of parsers and writers for the RDF/XML and
Turtle serializations of the RDF data model, and an RDF-storage module
that is written in C and designed to be tightly connected to Prolog.
The storage module provides fully-indexed lookup, statistics to support
a query optimizer, reliable persistent storage, transaction management
and full-text search.

    \item[JSON] (see also \secref{nqueens})
JavaScript Object Notation is a popular serialization format for
structured data.

    \item[HTTP] (SWI-Prolog and the web, \citeNP{wielemaker:tplp2008})
Supports both clients and servers.  We currently see the HTTP server
as one of the fundamental libraries.  See \secref{web} for an example
server.

    \item[ODBC]
Provides low-level access to ODBC databases. SWI-Prolog still lacks
high-level support, such as the one described by \citeN{Draxler:ALPUK91}.

    \item[TCP, UDP, SSL, TIPC]
These libraries provide basic network communication.

    \item[XPCE \cite{DBLP:conf/lpe/WielemakerA02}]
XPCE provides native and portable (X11 and MS-Windows) graphics.  As
described in the introduction, XPCE was part of SWI-Prolog from the
beginning. It is still the basis for many applications, including
the development tools described in \secref{environment}.
\end{description}

\section{Prolog as a web server}
\label{sec:web}

The web has become our most important application domain for SWI-Prolog.
A significant part of the interfaces described in \secref{interfaces}
have been influenced by the development of ClioPatria
(\textit{Thesaurus-based search in large heterogeneous collections},
\citeNP{Wielemaker:08a}). The user interface of ClioPatria is based on
AJAX using the YUI Widget set. In \secref{nqueens} we describe the
implementation of a small interactive web application that uses AJAX/YUI. We present our experiences with hosting
SWI-Prolog in Prolog in \secref{swiplorg}

\subsection{AJAX N-Queens -- how to web-enable Prolog}
\label{sec:nqueens}

When trying to adapt Prolog to the functioning of the Web, we encounter
what is often referred to as an ``impedance mismatch problem'': Prolog
is relational in that a query may map to more than one result, but HTTP
is essentially functional in that one query/request should map to
exactly one result/response. Sometimes this can be solved by using
\index{findall/3}\predref{findall}{3}, but this only works for a finite number of solutions and only
if there are not too many. Besides, we may \emph{prefer} to generate and
present the solutions ``a-tuple-at-a-time'', sometimes because it is
much cheaper in terms of memory requirements on both server and client,
and sometimes because we want to be able to decide, after having seen
the first couple of solutions, whether we want to see more.

Instead of wrapping queries in \index{findall/3}\predref{findall}{3} we choose to work with a
\jargon{virtual index} to the solutions that a query has. Each
solution in the sequence of $m$ solutions to a query receives an integer
index in the range $1..m$. This makes a query for the $i$-th solution of
a goal functional, and thus solves the impedance mismatch problem. So to
retrieve the first two solutions to an N-Queens solver, i.e., to do what
corresponds to the command-line session

\begin{code}
?- queens(8, L).
L = [1, 5, 8, 6, 3, 7, 2, 4] ;
L = [1, 6, 8, 3, 7, 4, 2, 5] .
?-
\end{code}

\noindent
we need to make \emph{two} HTTP requests: \texttt{/queens?n=8\&i=1}
followed by \texttt{/queens?n=8\&i=2}. Implementing this API in standard
Prolog is trivial if each HTTP request runs the query and returns the
$i$-th solution. In order to make this \emph{efficient}, we must make
sure that the system remembers the state. We achieve this by using a
Prolog thread (see \secref{threads}) and message queues to communicate
between the HTTP server threads and the \jargon{solver} threads. We use
a high-level abstraction for creating efficient a-tuple-at-a-time web
APIs to Prolog and programs written in Prolog:

\begin{description}
\predicate{thread_call}{6}{+ID, :Goal, +I, +Bindings, -Result, +Options}
Computes the \arg{I}-th solution to the (possibly) nondeterministic
\arg{Goal} in a thread uniquely identified as \arg{ID}. Succeeds exactly
once and binds \arg{Result} to a list of \mbox{\arg{Name}=\arg{Value}}
pairs that provides information about the result, including the bindings
specified in \arg{Bindings} if \arg{Goal} succeeded.
\end{description}

The idea is that when a call to \index{thread_call/6}\predref{thread_call}{6} exits, the thread
referenced by \arg{ID} may still be available and be ready to backtrack
and compute more solutions for indices \emph{greater} than \arg{I}. For
indices \emph{smaller} than \arg{I}, \index{thread_call/6}\predref{thread_call}{6} restarts the
enumeration of solutions. Below is an example session. \texttt{new=
\index{@false}\objectname{false}} indicates that the second call reuses a state.

\begin{code}
?- thread_call(t1, queens(8, L), 1, ['L'=L], R, []).
R = [success= @true, error= @false, errormessage= @null,
     bindings=['L'=[1, 5|...]], more= @true, new= @true, time=0.0].

?- thread_call(t1, queens(8, L), 2, ['L'=L], R, []).
R = [success= @true, error= @false, errormessage= @null,
     bindings=['L'=[1, 6|...]], more= @true, new= @false, time=0.01].
\end{code}

\noindent
Building an N-Queens web application server is just a matter of setting up
an HTTP server and declaring a handler which calls \index{thread_call/6}\predref{thread_call}{6}
and outputs the result as JSON. The code for the handler is shown
below:

\begin{code}
:- http_handler(root(queens), queens, []).

queens(Request) :-
    http_parameters(Request, [n(N,[integer]), i(I,[integer])]),
    http_session_id(ThreadID),
    thread_call(ThreadID, queens(N, L), I, [queens=L], Result, []),
    term_to_json(Result, JsonTerm),
    reply_json(JsonTerm).
\end{code}

\noindent
The module \texttt{term_to_json} provides the means to convert any Prolog
term into a JSON structure. Above it is used for converting
the result from a call to \index{thread_call/6}\predref{thread_call}{6} into a JSON term which is
then written to output using \index{reply_json/1}\predref{reply_json}{1} from library
\file{http/http_json}. For example, the HTTP request
\texttt{/queens?n=8\&i=4} results in the response depicted below:

\begin{code}
{ "success":true, "more":true, "error":false, "errormessage":null,
  "bindings": {"queens": [1,7,5,8,2,4,6,3]}, "new":false, "time":0.01
}
\end{code}

\noindent
On the client side, nothing is new or in any way peculiar to the use of
Prolog. Thanks to the pure JSON interface provided, the client side is a
just an ordinary AJAX application that can be written by any programmer
familiar with the languages and techniques involved. For our demo, we
chose to work with YUI, but we could just as well have used any of the
many alternatives available. The GUI to our N-Queens demo is shown in
\figref{nqueens}.

\postscriptfig[width=0.5\linewidth]{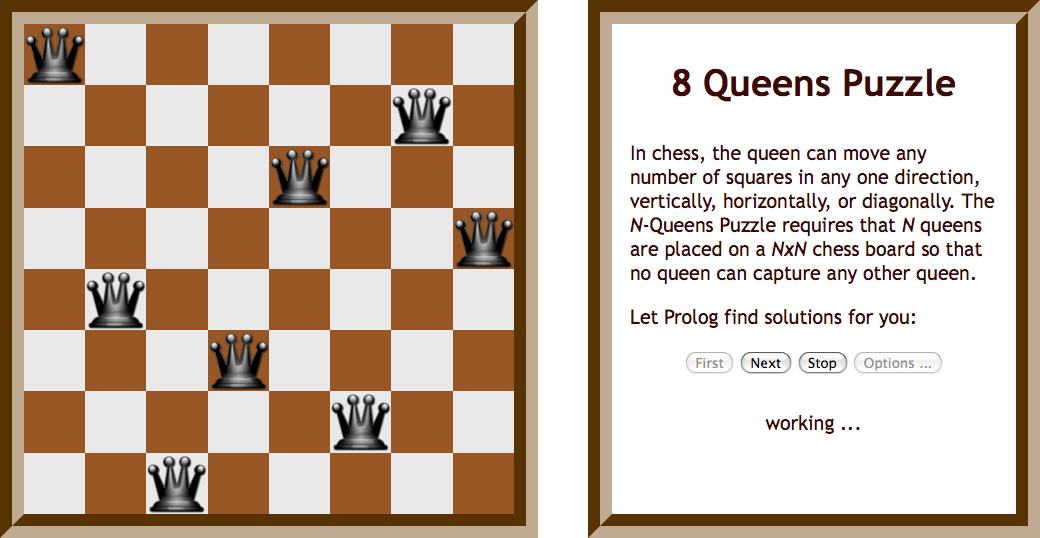}{The N-Queens demo}

\subsection{The SWI-Prolog website}
\label{sec:swiplorg}

Since February 2009, SWI-Prolog's website is implemented using the
SWI-Prolog HTTP server library. The basis is formed by PlDoc
\cite{Wielemaker:2007c}, the SWI-Prolog literate programming system that
provides a web-interface for the documentation of loaded code and the
system manuals. SWI-Prolog hosting itself has two advantages: (1) it
provides a realistic environment for testing the HTTP libraries and (2)
PlDoc provides a uniform interface for all documentation and
automatically creates links from documents in Wiki format to the
documentation. \Tabref{traffic} gives some statistics on the traffic
that is handled by the site.

\begin{table}
\begin{tabular}{ll}
\hline
Machine & $2 \times$ AMD Opteron 2356 (quad core), 32 Gb memory \\
\hline
Hits per day & 22,117 \\
Visits per day & 2,759 \\
Traffic per day & 3,376 Mb \\
CPU per day & 1,632 seconds \\
Memory usage & 25 Mb \\
\hline
\end{tabular}
    \caption{Average daily traffic from \protect\url{www.swi-prolog.org},
	     September 2009.}
    \label{tab:traffic}
\end{table}

\paragraph{Provided services}

The server provides several different services: dynamically generated
pages (serving from PlDoc and Wiki pages), small static pages (css,
JavaScript, images, etc.), large static pages (downloads of binaries,
sources and PDF documentation), and CGI for supporting gitweb, the web
frontend of the GIT source-code management system.

\paragraph{Stability and scalability}

The server uses the core parts of the SWI-Prolog (web)infrastructure:
the multi-threaded HTTP libraries and XML parsing.  Running this server
$24\times{}7$ on the web has revealed four critical bugs (three of which
were related to concurrency) and several memory leaks.  Currently, the
server runs without problems.
\section{Language properties and extensions}
\label{sec:language}

SWI-Prolog is first of all a system for prototyping medium-scale (50-100
K-lines) applications, where Prolog is used as glue to unite external
resources such as graphical libraries and (RDF) datastores. This idea has
shaped the implementation.

\subsection{The compiler and program loading}
\label{sec:compiler}

The compiler is distantly based on the ZIP abstract machine
\cite{Bowen:83,Neumerkel:93}. The SWI-Prolog VM is a structure copying
machine that passes arguments through the environment and addresses
arguments using an \jargon{argument pointer}. The argument pointer is
also used to read and write arguments in lists and compound terms. The
C-based emulator currently implements 145 instructions. The
garbage collector is a mark-and-sweep collector that closely follows
\citeN{carlson:88}.

While extending the instruction-set, we ensured that both compilation
and decompilation remain simple tasks. The compiler is written in C and
used both for compiling source files and adding dynamic code using
\index{assert/1}\predref{assert}{1}. As a result, static code and dynamic code have the same
reflexive properties (e.g., \index{clause/2}\predref{clause}{2} can be used for both).

Our VM allocates all variables on the stack and passes all arguments
over the stack. Where the WAM loses access to arguments and temporary
variables (variables that only appear in two adjacent calls in the
body), we allocate all these variables. An advantage of this is that the
source-level tracer can display the value of all variables in a clause
(see \figref{guitracer}). As a consequence, however, our environments
are larger and without precautions, more data remains accessible through
the environments. In \citeN{CYCLOPS/Wielemaker/2008} we show that the
reachability problem can be solved at marginal costs by scanning the
virtual machine code to determine which variables are initialized and
still reachable.

The time needed for loading a large program (and for updating it after
some of its source files are modified) must be short, lest the process
of developing a prototype become too cumbersome. In particular, we are
interested in two features: (1) it must be possible to compile files
clause-by-clause, so there is no need for buffering, and (2) it must be
possible to make code available \jargon{on-demand}, so there is no need
to compile the whole program before execution can start. Starting with
version 5.7, this is realized by the predicate \jargon{supervisor-code}.
A predicate contains a linked list of clauses compiled to ZIP VM code
and a supervisor. Execution of a predicate starts at the entry-point of
the supervisor. Every predicate has the same initial supervisor,
consisting of the single instruction \const{S_VIRGIN}. If this
instruction is executed, it searches for the implementation of the
predicate using this sequence:

\begin{enumerate}
    \item If already defined, we are done.
    \item If the predicate is in one of the \jargon{import}
    modules (see \secref{importmodule}), import it.
    \item If the predicate is in the auto-load (lazy load)
    index, compile the source.
    \item If the predicate is still undefined, replace the supervisor
    with \const{S_UNDEF}, an instruction that handles calls to
    undefined predicates depending on the \const{unknown} flag.
\end{enumerate}

After establishing the clauses for the predicate, \const{S_VIRGIN}
examines the clause-heads by decompilation and generates the
clause-indexing code.

Note that the supervisor only deals with lazy loading, lazy generation
of indices, calling non-Prolog (i.e., foreign) predicates and
clause-\emph{selection}. In particular, it carries no information that
is not available in the predicate attributes and clause-list. This
approach allows for \jargon{hotspot} compilation and other optimizations
without complicating the reflexive features of SWI-Prolog
\cite{DBLP:conf/iclp/SilvaC07}.

\subsection{The module system}

Compatibility requirements during early development (see
\secref{intro}) have caused SWI-Prolog to adopt the Quintus Prolog
predicate-based module system. However, we have made several
modifications to this model to make it more suitable for rapid
development and lazy loading of code. Using modules in Prolog has great
practical value for two reasons: modules help avoid name-conflicts,
especially for local \jargon{helper} predicates, and they define the
public interface of a file. This section describes and motivates the
modifications we made to the original Quintus Prolog model.

\subsubsection{Meta-predicate handling}

Meta-predicates are predicates that refer to other predicates. For
example, \index{findall/3}\predref{findall}{3} takes a \jargon{goal} as argument. With a module
system there can be multiple predicates with the same name and arity, and
a predicate must refer to the correct one: the one that appears in the
same lexical context. For example, given the following program,
\index{findall/3}\predref{findall}{3} (which is defined in the \const{system} module) must call
\index{child/2}\predref{child}{2} in the module \const{family}.

\begin{code}
:- module(family, [ child/2, children/2 ]).

child(bob, jane).
child(bob, peter).

children(Parent, Children) :- findall(Child, child(Parent,Child), Children).
\end{code}

\noindent
Predicate-based module systems solve this problem by declaring \index{findall/3}\predref{findall}{3}
as a meta-predicate: \verb$:- meta_predicate findall(?,0,-).$ An
argument that is passed module-sensitive information (e.g., a goal) is
specified by `:' or by an integer. An integer specifies that the
argument is a goal and this goal will be called with $N$ additional
arguments (e.g., \term{maplist}{2,?,?}). This declaration is
processed when the compiler compiles \emph{a call to} a meta-predicate
and causes the compiler to embed the argument in a term \bnfmeta{module}:\bnfmeta{plain}
(i.e., to \jargon{qualify} the argument;
\verb$family:child(Parent,Child)$ in the example above).

This approach is used by many Prolog systems, but it comes with two
drawbacks: (1) the compiler must have access to information about
whether any given predicate is a meta-predicate, and (2) modifying the
meta-predicate declarations requires all code that \emph{calls} this
predicate to be recompiled. The first requirement either puts ordering
constraints on the location of meta-predicate declarations or requires
multi-pass compilation. With lazy loading, the index of predicates that
can be loaded must be known at compile-time and must include
meta-argument information. The second requirement complicates
resynchronizing the Prolog database if one or more source file has
changed (see the discussion of \index{make/0}\predref{make}{0} in \secref{editcycle}).

SWI-Prolog supports the \index{meta_predicate/1}\predref{meta_predicate}{1} directive without changing the
compilation of code \emph{calling} a meta-predicate. This is achieved by
adding a \jargon{context module} to each environment. If an environment
is created, the context is copied from the parent. Next, the virtual
machine resolves the predicate (possibly through lazy loading). If the
predicate is not a meta-predicate the context is set to the module in
which the predicate is defined. Next, the virtual machine starts
executing the supervisor (see \secref{compiler}) of the predicate. The
supervisor of a meta-predicate qualifies all meta-arguments and then
sets the context to the module of the predicate.

This implementation satisfies our requirement (of being able to autoload
meta-predicates and update meta-declarations dynamically) at the cost of
some space and runtime overhead in managing the context module in each
environment.

\vspace{-2mm}
\subsubsection{Import modules}
\label{sec:importmodule}

The notion of import modules generalizes the distinction between
built-in and user-defined predicates found in other Prolog systems. If a
predicate is not locally defined, the system first tries to import it
silently from the modules' import modules. In the normal setup, each
user module imports from the \const{user} module, which in turn imports
from \const{system}. The \const{system} module contains the built-in
predicates. The underlying machinery allows for multiple import modules
per module and an arbitrary acyclic module-dependency graph. This
mechanism is used to create unit-tests as isolated modules importing
from the module-to-be-tested in PlUnit (see \secref{plunit}).

Especially for rapid development, programmers may choose to import
utilities that are used at many places in an application into the
\const{user} module. This makes these utilities available from the
top level for debugging and avoids the need to import them in every
application module. Note that, while a definition that is visible in the
\const{user} module \emph{can} be used in a module without explicit
import, it is still \emph{allowed} to import explicitly.

Import modules allow for different reuse schemes. SWI-Prolog supports
Prolog de-facto standard import the predicates \index{use_module/[1,2]}\predref{use_module}{[1,2]}. It
can support modularity similar to C by loading all modules into module
\const{user} and omitting explicit import relations between the modules.
Finally, it supports modules that inherit from their context (as used in
PlUnit). Future versions are likely to split the \const{system} module
into multiple modules to accommodate subsystems such as \const{iso}.

\subsubsection{Operator handling}

Most Prolog systems, even those that provide modules, use globally
scoped operators. However, integration of large programs that feature
programmer-defined operators is likely to fail due to operator
conflicts. In the most common case, this results in syntax errors. In
other cases it results in different interpretation of terms that cause
different behavior of the program. Such cases are hard to diagnose.

Therefore, we decided to make operators local to the module in which it
is declared. The system searches for operators in the same order as it
searches for predicates (see \secref{importmodule}). This scheme allows
for defining globally used operators in the \const{user} module. Support
for global operators is needed for compatibility. Operators can be
exported and imported. Here is an example from the library
\file{record.pl}, which provides named access to fields in structures.

\begin{code}
:- module((record), [ (record)/1, op(1150, fx, record) ]).
\end{code}

\noindent
\subsection{Avoiding limitations}

We have attempted to avoid hard limits that could complicate the task of
writing an application. In particular:

\begin{description}
\item[Atoms]
There is no limit on the length of atom names, which can be written in
Unicode and include null characters. If atoms are used for processing
input (text) this is needed to avoid representation errors, either on
the length or the represented characters. Allowing for null characters
allows representing arbitrary data (e.g., \jargon{image data}) as atoms.
Processing input using atoms calls for atom garbage collection.

\item[Integers]
Having no bounds to integers is not only meaningful to mathematicians.
It also allows representing integers in, e.g., XML documents as Prolog
integers, without worrying about overflows. When building interfaces to
foreign resources, it covers all limited integer types in a clean and
uniform way.

\item[Terms]
Compound terms have unbounded arity, which makes them particularly
suitable for implementing arrays. SWI-Prolog supports rational trees
(also called cyclic terms). Although we are not convinced that rational
trees have much practical value in Prolog, crashing or looping on them
is not acceptable. This is particularly the case for servers (DoS
attacks) and education because students frequently create rational trees
unintentionally (e.g., \verb$List = [Head|List]$). SWI-Prolog can run in
three modes, causing the unification above to succeed with a rational
tree (default), to fail or to throw an error, depending on the Prolog
flag \const{occurs_check}.

\item[Stacks]
Unfortunately, SWI-Prolog's stacks are limited to 128Mb per stack on
32-bit hardware.  Given that 64-bit systems are now widely available we
do not plan to raise the limits on 32-bit hardware.
\end{description}

\subsection{Global variables and destructive assignment}

A global variable associates an identifier (we only allow atoms)
with a term on the heap.  We provide two types of assignment to global
variables: backtrackable (\index{setval/2}\predref{setval}{2}) and non-backtrackable (\index{nb_setval/2}\predref{nb_setval}{2}).
The naming and implementation is based on hProlog \cite{Demoen:CW350}.

Assigning a value to a global variable is a destructive operation. The
same implementation can be used to facilitate destructive assignment of
arguments of compound terms. Global variables and destructive
modification of compound terms are useful in combination, for example to
implement a global array, as shown below:

\begin{code}
new_global_array(Name, Size) :-
        functor(Array, array, Size),
        setval(Name, Array).

global_array_set_element(Index, Name, Value) :-
        getval(Name, Array),
        setarg(Index, Array, Value).
\end{code}

\noindent
The backtrackable \index{setarg/3}\predref{setarg}{3} is supported by many Prolog implementations.
Non-backtrackable assignment as implemented in \index{nb_setarg/3}\predref{nb_setarg}{3} is less
widely supported. GNU-Prolog supports it using \index{setarg/4}\predref{setarg}{4}, but the
argument value must be atomic. Backtrackable assignment is based on
two-cell entries in the trail that maintain the old value.

Non-backtrackable assignment of a value that lives on the heap is more
complicated. It is achieved by maintaining a global pointer (called
\jargon{frozen_bar}) to the top of the heap at the moment of assignment.
Backtracking never resets the top-of-heap below this mark. This implies
that data that is older than the global term must be discarded by the
garbage collector instead of by resetting the top-of-heap in
backtracking. Backtracking may reset trailed bindings inside the value
if the value is compound. This is indeed the case in hProlog. SWI-Prolog
avoids this by making a copy of the value if the value is compound.

Sometimes, it is necessary to preserve state over backtracking. A clean
solution to that are the all-solution predicates (e.g., \index{findall/3}\predref{findall}{3}). In
standard Prolog, one can only use dynamic predicates if the all-solution
predicates are not appropriate. The global nature of dynamic predicates
make it hard to implement reentrance, thread-safety and clean up in the
event of an exception. One solution to this problem is given by
\citeN{DBLP:journals/corr/abs-0808-0556}, by introducing explicit
interaction with Prolog engines. We support this style of programming
by using threads. Non-backtrackable assignment in compound terms
provides another solution. The example below counts proofs for a goal.
It is fast, safe and runs in constant space.

\begin{code}
proof_count(Goal, Count) :-
        State = count(0),
        (   Goal,
            arg(1, State, C0), C1 is C0 + 1, nb_setarg(1, State, C1),
            fail
        ;   arg(1, State, Count)
        ).
\end{code}

\noindent
\subsection{Attributed variables and coroutining}
\label{sec:language:attvars}

\jargon{Attributed variables} \cite{christian:attvars} were added in
early 2004 to allow for constraints and coroutining. For SWI-Prolog we
chose to use the dynamic interface for attributed variables that was
developed by Bart Demoen \cite{Demoen:CW350}. This interface does not
require attributes to be declared, and represents them with a linked
list associated with the variable. This interface (which is currently
available in hProlog, SWI-Prolog, XSB, Ciao, YAP and SICStus) is the
basis of the portable constraint libraries discussed in \secref{clp}.
Attributed variables are also used to implement the common coroutining
predicates: \index{freeze/2}\predref{freeze}{2}, \index{when/2}\predref{when}{2} and \index{dif/2}\predref{dif}{2}.

\subsection{Multi-threading}
\label{sec:threads}

Multi-threading was initially introduced to support scalable web servers
in Prolog. The design and implementation is described in
\citeN{Wielemaker:03c} and is the basis for the ISO WG17 work on
threading in Prolog as well as multi-threading support in YAP and XSB.
In our design, each thread comes with an independent set of stacks. This
implies that threads cannot share terms and therefore unification,
backtracking and garbage collection in each thread can be done
independently from other threads. The key features are:

\begin{itemize}
    \item Static predicates are fully shared.  SWI-Prolog provides
    both shared and non-shared dynamic predicates.\footnote{In
    SWI-Prolog, the default is to provide shared dynamic predicates.
    In XSB, dynamic predicates are by default non-shared.  This is
    a safer choice because sharing dynamic data almost always
    calls for additional synchronization.}

    \item Communication is primarily achieved by using message queues
    (\jargon{ports}) holding Prolog terms.

    \item There is a signaling interface that allows a thread to interrupt
    the execution of another thread. This interface was initially
    intended to support the debugger, but is also used to abort threads
    by injecting an exception into their control flow.

    \item Threads are layered on top of the POSIX thread primitives,
    providing smooth integration with thread-safe foreign code and
    taking full advantage of multi-core hardware.
\end{itemize}

We see two major application areas for concurrency in Prolog: (1)
solving a large problem, and (2) solving many, mostly independent,
tasks. Supporting problems of the first class is problematic because
such problems often require intensive communication for which copying
terms is too expensive or for which the copying semantics is
inappropriate. Another problem is that when a thread encounters a
serious error (e.g., a resource error), this may affect the entire
computation (which may have to take appropriate action).

The design is successfully applied for the second class of problems,
dealing with many, mostly independent, tasks. The SWI-Prolog HTTP
libraries (\secref{web}) have grown into a mature multi-threaded
web server.

\subsection{The C-interface}
\label{sec:capi}

As we have seen in the introduction, the C-interface was one of the
success factors of SWI-Prolog in its early days. Mutually recursive
calling between C and Prolog is now commonly supported. Below we
describe the more distinguishing features of the current C-API.

\paragraph{Non-determinism}

C-API supports non-deterministic foreign predicates by adding a context
argument that provides the type of call. We distinguish three reasons to
call the C function:

    \begin{description}
        \item[\const{PL_FIRST_CALL}]
	The first call is the same as for deterministic foreign code.
	The function can return with one of three values: \const{FALSE}
	to indicate failure; \const{TRUE} to indicate deterministic
	success or \const{RETRY} with an integer or pointer context. The
	context must carry enough information to compute the next
	solution. For example, when enumerating values from an array,
	this could be the array index. Often, the foreign predicate
	allocates a context structure and returns a pointer
	to this structure.

	\item[\const{PL_REDO}]
	This call is issued by the kernel upon backtracking to an
	invocation of a foreign predicate. All Prolog arguments are
        guaranteed to be the same as for the initial goal. The
	implementation uses the previously returned context to compute
	the next result and returns with one of the three values, just
	as with the \const{PL_FIRST_CALL} call.

	\item[\const{PL_PRUNED}]
	Called if the choice-point is pruned as a result of executing
	a cut or handling an exception. Only the context value is valid.
	The foreign implementation must clean up side-effects (e.g.,
	free memory allocated for preserving the context).
    \end{description}

\paragraph{Foreign context frames}

The C-API refers to Prolog terms through explicitly allocated
\jargon{term-handles}. Because Prolog keeps track of the allocated
handles, it knows which terms are references from C-code and can perform
heap and atom garbage collection transparently. This mechanism was first
introduced by Quintus Prolog and is now used in many modern Prolog
implementations. Term-handles that are used in the implementation of
predicates in C are discarded when the C-implementation returns to
Prolog. Quintus does not provide an API to deallocate term-handles.

Setting up a call from C to Prolog involves allocating term-handles for
constructing the arguments and processing the resulting bindings. If the
overall application control is in C and the C-code makes multiple calls
to Prolog, we need some way to discard term-handles. SWI-Prolog
implements this by means of \funcref{PL_open_foreign_frame}{} \ldots
\funcref{PL_close_foreign_frame}{}. All term-handles created between these two
matching calls are invalidated.\footnote{SWI-Prolog also provides
\funcref{PL_reset_term_refs}{} to discard the argument-handle and all term-handles
created afterwards. This function was copied by SICStus as
\funcref{SP_reset_term_refs}{} when porting XPCE to SICStus.} In addition,
\funcref{PL_rewind_foreign_frame}{} rewinds (i.e., backtracks) the heap to the
state at \funcref{PL_open_foreign_frame}{}. Rewinding can be used to try an
alternative if a sequence of \funcref{PL_unify_*}{} calls fails (see
next bullet).

\paragraph{Hand-crafted wrappers}

The Quintus C-API is designed to describe the C-code from Prolog and
automatically generate a wrapper for it. The generated API is typically
not Prolog-friendly. Values returned by C functions must be carefully mapped to Prolog
success/failure or an exception. Using the Quintus approach the final
mapping to a natural Prolog API must be done in Prolog. This is
particularly cumbersome when dealing with enum types or \verb$#define$
constants.

The SWI-Prolog C-API concentrates on passing Prolog terms and supporting
Prolog success, failure and errors. This implies that the
\jargon{wrapper} is hand-crafted\footnote{The library qpforeign.pl
provides a Quintus-compatible wrapper generator} and therefore we
provided a third family of functions: PL_unify_\bnfmeta{type}(Term, C-Value),
which unifies a Prolog argument with a converted C-value and returns
\const{TRUE} or \const{FALSE}.

We illustrate our approach in \figref{getenv}. The actual wrapper is the
function \funcref{pl_getenv}{}. The first call extracts the name of the requested
variable if it is an atom. The text is extracted as a C-string in the
native locale.\footnote{Internally all text is stored as Unicode.} If
the extraction fails, it leaves an appropriate exception in the
environment. If the \funcref{getenv}{} API fails, we raise an
exception.\footnote{We are planning to provide the functionality of
\funcref{existence_error}{} in the Prolog C-API.} Finally, if all went well, we unify the extracted
string with the second argument and return the success of this
unification. The module can be compiled, loaded and used as shown in the
example below. The \program{swipl-ld} utility is a wrapper around the C
compiler that hides platform-specific details.

\begin{code}
1 ?- load_foreign_library(getenv).
true.
2 ?- my_getenv('HOME', X).
X = '/home/janw'.
3 ?- my_getenv(notavar, X).
ERROR: environment_variable `notavar' does not exist
\end{code}

\noindent
\begin{figure}[ht]
\begin{code}
#include <SWI-Prolog.h>
#define MKFUNCTOR(name, arity) PL_new_functor(PL_new_atom(name), arity)
static functor_t FUNCTOR_error2, FUNCTOR_existence_error2;

static int existence_error(term_t missing, const char *what)
{ term_t ex;
  if ( (ex = PL_new_term_ref()) &&
       PL_unify_term(ex, PL_FUNCTOR, FUNCTOR_error2,
                           PL_FUNCTOR, FUNCTOR_existence_error2,
                             PL_CHARS, what, PL_TERM, missing,
                           PL_VARIABLE) )
    return PL_raise_exception(ex);
  return FALSE;
}

static foreign_t pl_getenv(term_t name, term_t value)
{ char *ns, *vs;
  if ( !PL_get_chars(name, &ns, CVT_ATOM|CVT_EXCEPTION|REP_MB) )
    return FALSE;
  if ( !(vs=getenv(ns)) )
    return existence_error(name, "environment_variable");
  return PL_unify_chars(value, PL_ATOM|REP_MB, -1, vs);
}

install_t install_getenv()
{ FUNCTOR_error2           = MKFUNCTOR("error", 2);
  FUNCTOR_existence_error2 = MKFUNCTOR("existence_error", 2);
  PL_register_foreign("my_getenv", 2, pl_getenv, 0);
}
\end{code}

\noindent
    \caption{Foreign wrapper for \funcref{getenv}{}}
    \label{fig:getenv}
\end{figure}
\section{The development model for SWI-Prolog}
\label{sec:testing}

A language like SWI-Prolog is only taken seriously by its users if
its implementation is stable and dependable. This desire for stability
is at odds with the academic desire for exploration and development of
new language features. In order to reconcile these two desires, the
development of SWI-Prolog proceeds in branches according to well-known
Open Source practices described in \citeN{bazaar}. In particular:

\begin{itemize}
    \item Even-numbered branches are stable and odd-numbered
          branches are for development.

    \item Create a new branch when 1) the
          development branch appears to be stable (i.e.,
	  there are few reported stability issues), and 2)
	  new developments are about to aversely impact stability.
	  The stable branch is supposed to provide a stable core,
	  but newer libraries may be unstable.

    \item Release often.  The typical release cycle on the
          active development branch is 2 weeks, but can be
	  shorter if the users require particular fixes or
	  functionality.  It can be longer if the current
	  state is considered too unstable or there are no
	  changes of interest.

    \item Respond quickly.  When possible, provide
          a fix or work-around for problems communicated on the
	  mailing list or posted on the bug-tracking system.
	  If the problem is expected to affect many users, make a
	  release. Otherwise, the patch is only available through the
	  GIT\fnurl{http://git-scm.com/} repository. Professional users
	  are expected to be able to build the system from the GIT
	  distribution.

    \item Maintain a regression test suite and run it frequently.
          We do not have the resources to create a comprehensive
	  test suite. As an in-between solution, we typically create
	  test-cases from bug-reports.  This scheme avoids bugs from
	  reappearing.
\end{itemize}

\subsection{The regression test suite}
\label{sec:testsuite}

The regression test suite is activated through the GNU Makefile
standard target \const{check}. For historical reasons, the
tests are built according to two different test paradigms:

\begin{itemize}
    \item Single-clause tests that are expected to succeed.  The predicate name
    	  indicates the set of tests and the first argument is a ground
	  term that identifies the test.  The test driver enumerates the
	  tests using \index{clause/2}\predref{clause}{2}.

    \item Scripts.  A script is a Prolog file, typically named
    	  test_\bnfmeta{topic}.pl. It defines a module test_\bnfmeta{topic} that exports
	  test_\bnfmeta{topic}/0. The test driver enumerates all files in a
	  directory, loads them and runs the exported goals. Scripts
	  are used to test larger programs (e.g., solve a constraint problem).
	  More recently, they also contain unit-test suites as described in
	  \secref{plunit}.
\end{itemize}

Currently, the test suite for the core system contains 411 tests of the
first form and 58 test scripts. 35 of the 58 test scripts are
\textit{PlUnit} suites, providing another 353 individual tests.

\subsection{PlUnit: the test driver framework}
\label{sec:plunit}

\textit{PlUnit}\fnurl{http://www.swi-prolog.org/pldoc/package/plunit.html}
is the test driver framework of SWI-Prolog. Unlike the driver framework
outlined above, \textit{PlUnit} is targeted at users of SWI-Prolog. It
is based on the idea of single-clause tests, but uses a slightly
different way to identify test-clauses and adds a second argument that
specifies properties of the body, such as expected bindings,
non-determinism, failure, or raised exception. The second argument can
also specify a condition to run the test, and a goal to setup and clean
up the environment to run the test. Below is a simple example:

\begin{code}
:- begin_tests(aggregate).
test(aggregate_count, Count == 2) :-
        aggregate(count, X^between(1,2,X), Count).
:- end_tests(aggregate).
\end{code}

\noindent
A description of the expected behavior of the body allows the test
driver to give a more descriptive report if a test fails (contrasting
the expected and actual behavior).

A test-block (\const{begin_tests}..\const{end_tests}) is compiled to a
module that inherits from its lexical context (see
\secref{importmodule}). This allows test units to be embedded in the
actual source code: the tests have access to the internals of the
module to be tested but do not pollute the namespace of this module. The
\textit{PlUnit} driver can be asked whether or not embedded tests should
be compiled and whether or not they should be run automatically by
\index{make/0}\predref{make}{0} (see \secref{editcycle}) after the module has been modified.
\section{Discussion and future work}
\label{sec:future}

SWI-Prolog has become a comprehensive and mature implementation of the
Prolog language. Its focus is on integrating technology from the logic
programming community and interfacing to external resources to provide a
platform for prototyping and development of fairly large applications.
The system is widely used in educational, research and commercial
environments.

Prolog still has a difficult marketing position. It is generally
perceived as hard-to-learn, lacking ready-to-use resources and a good
Integrated Development Environment (IDE). Nevertheless, the Prolog
language is being used in new projects. In such projects, we typically
find a mixture of four components: (1) application-specific high level
languages, and (2) rule-based reasoning, and (3) constraint handling,
and (4) Semantic web (RDF) data.

Prolog is well equipped to compile high-level application-specific
descriptions into programs that combine the other three components.
Using application-specific descriptions is particularly suitable for
domains that face frequent changes in the rules and
procedures. For example,
SecuritEase\footnote{\url{http://www.securitease.co.nz/}} uses CHR for
transforming their Constraint Query Language into SQL. Although we lack
hard evidence, we think that many commercial users deploy SWI-Prolog as
a multi-threaded server component.

In many universities, Prolog is now taught as part of a course on
programming paradigms. After a few weeks, students often come to the
conclusion that logic programming is a neat idea, but it is not useful
for anything practical. Possibly, Prolog should be taught in the context
of, e.g., the Semantic Web, where students can create applications using
real data that is readily available on the web, and can compare writing
Prolog rules over this data with querying this data in an
imperative language through a SPARQL interface.

For a long time, the Prolog community was divided into many isolated
islands. Adoption of the ISO-standard, although not perfect
\cite{DBLP:conf/iclp/SzaboS06}, and the development of larger portable
resources such as Logtalk, Leuven CHR, CLP(FD) and CLP(Q,R) have built
bridges between these islands. Developers of portable resources persuade
Prolog system developers to resolve incompatibilities. At the same time,
the existence of portable resources makes the logic programming
community more credible.

Future SWI-Prolog development will concentrate on the following aspects:

\begin{itemize}
    \item Improving compatibility, notably with systems with a similar
	  module system.
    \item Improving stability, scalability and performance.
    \item Improving support for rule-based programming by providing
          tabling.
    \item Providing more libraries, notable for RDF and web programming.
    \item Improving the development environment, notably by adding a
          type-checker \cite{DBLP:conf/iclp/SchrijversCWD08}, adding
	  (style) warnings, and adding tools that support
          refactoring of programs.
\end{itemize}

\section*{Acknowledgements}

Continuous development of SWI-Prolog has been made possible by the SWI
department of the University of Amsterdam. Currently, the development is
hosted at the web and media group of the VU University Amsterdam.

An important factor in the success of SWI-Prolog is the community that
provides the motivation, challenges and bug reports, as well as code,
patches and descriptions of how to fix errors in the implementation. In
particular, we would like to acknowledge Paulo Moura for Logtalk, his
effort in pointing out incompatibilities between Prolog implementations
and his work on the MacOS port; Richard O'Keefe and Bart Demoen for
background knowledge of libraries, standards, and implementation details,
and Ulrich Neumerkel for testing and code.

\bibliography{pl,phd,sw,clpfd,chr,wlpe03}

\end{document}